# Physics design for the C-ADS main linac based on two different injector design schemes*

YAN Fang[1] (闫芳), LI Zhi-Hui（李智慧）, MENG Cai（孟才）, TANG Jing-Yu（唐靖宇）SUN Biao (孙彪)

Institute of High Energy Physics, Chinese Academy of Science, Beijing, 100049, China

**Abstract:** The China ADS (C-ADS) project is proposed to build a 1000 MW Accelerator Driven sub-critical System around 2032. The accelerator will work in CW mode with 10 mA in beam current and 1.5 GeV in final beam energy. The linac is composed of two major sections: the injector section and the main linac section. There are two different schemes for the injector section. The Injector-I scheme is based on a 325-MHz RFQ and superconducting spoke cavities of same RF frequency and the Injector-II scheme is based on a 162.5-MHz RFQ and superconducting HWR cavities of same frequency. The main linac design will be different for different injector choice. The two different designs for the main linac have been studied according to the beam characteristics from the different injector schemes.

## 1 Introduction

The C-ADS project is proposed to build a 1000-MW Accelerator Driven sub-critical System around 2032. The driver accelerator will work in Continuous Wave (CW) mode, with the final energy of 1.5 GeV and average beam current of 10 mA. The C-ADS linac includes two major sections: the injector section and the main linac section. The injectors accelerate the proton up to 10 MeV and the main linac boost the energy from 10 MeV up to 1.5 GeV. The general layout of the linac is shown in Figure 1. To satisfy the restrict stability and reliability command [1] of C-ADS driver linac in lower energy part, there will be two identical Injectors operating paralleled backing up for each other. In the main linac part, the local compensation method is applied to achieve a fault-tolerance design which allows failures of key components such as cavities and focusing elements without beam interruption.

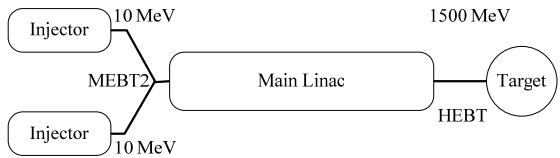

Figure 1: Layout of the C-ADS linac.

At present, two different design schemes for the injectors are proposed [2, 3], with Scheme I based on 325 MHz (the same frequency with main linac) and Scheme II based on 162.5 MHz. Finally only one scheme will be chosen and two identical injectors will be built and operates as a hot spare stand-by. For both design schemes, the injector is composed of an Electron Cyclotron Resonance (ECR) ion source, a Low Energy Beam Transport (LEBT) line, a Radio Frequency Quadrupole (RFQ), a Medium Energy Beam Transport (MEBT) line and a superconducting section. There will be a matching section – MEBT2 [4] to transfer the beam from any of the two injectors to the main linac. The beam parameters at the exit of the RFQ for the two schemes are listed in Table 1.

Table 1: Beam parameters at the exit of the RFQ

| Parameters | Scheme I | Scheme II |
|---|---|---|
| Frequency/MHz | 325.0 | 162.5 |
| $\varepsilon_t$/mm.mrad | 0.20 | 0.28/0.2626 |
| $\varepsilon_l$/mm.mrad | 0.17 | 0.288 |

The main difference of the two Injector schemes is the frequency choices. This will inflect the main linac design effectively. As the average current in the main linac will be the same, with half frequency, the Scheme II bunch current will be nearly doubled comparing with Scheme I. The different space charge effect caused different lattice structures for obtaining a stable beam dynamics design. Besides, for Scheme II the frequency jump in the front of the main linac will ask for a larger longitudinal acceptance and could cause potential troubles in longitudinal beam dynamics. The bigger emittance coming from Scheme II alleviates the space charge effect in one way, but in the mean while also requires a bigger acceptance in the main linac. Although the philosophy and criteria for the design of the main linac are the same with two injector schemes, the lattice structure will be different because of different beam characters. In this paper, the design details, considerations and the results of the two main linacs basing on two different injector schemes are described and presented.

## 2 General considerations on the main linac design

To fulfil the strict reliability constrains, over-design, redundancy and fault tolerance strategies are implemented in the basic design. To cover the whole energy range from 10 MeV to 1.5 GeV in the main linac section, we need at least four types of superconducting cavities. After optimization, we have chosen two single-spoke cavities working at 325 MHz with geometry betas of 0.21 and 0.40, respectively, and two 5-cell elliptical cavities working at 650 MHz with geometry betas of 0.63 and 0.82, respectively. The acceleration efficiencies of the four cavities and their effective energy ranges are shown in Figure 2. The effective energy ranges for the four types of cavities are all shifted to the lower energy to accommodate the special phase advance law required by the stable beam dynamics. The parameters of the cavities are listed in Table 2. For the nominal design, de-rated cavity voltage is used, and 30% (comparing to nominal design) cavity voltage is reserved for the local compensation. In the mean while this redundancy also benefits the cavity reliability.

*Supported by China ADS project (XDA03020000)
1) yanfang@ihep.ac.cn

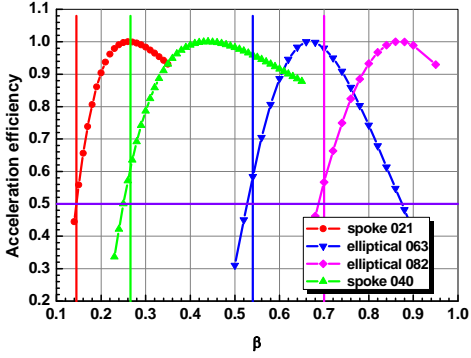

Figure 2: Acceleration efficiency of the cavies in main linac

Table 2: Parameters of the cavities in the main linac

| Cavity type | βg | Freq. MHz | Vmax MV | Emax MV/m | Bmax mT |
|---|---|---|---|---|---|
| S-Spoke | 0.21 | 325 | 1.64 | 31.14 | 65 |
| S-Spoke | 0.40 | 325 | 2.86 | 32.06 | 65 |
| 5-cell ellip. | 0.63 | 650 | 10.26 | 37.72 | 65 |
| 5-cell ellip. | 0.82 | 650 | 15.63 | 35.80 | 65 |

The lattice structures for each section of the main linac are shown in Figure 3, and they are characterized by long drift in both side of each period. An 800 mm spacing exists between periods for accommodating cryomodule from warm-to-cold transition. With this kind of lattice structure, the cryomodule structure is more flexible. It allows short cryomodule housing several periods or even one period without affecting the beam dynamics performance. The cryomodule can't be too long, on the one hand, the alignment for too many devices could be a problem, and on the other hand warm beam dynamics devices are needed between the periods for beam tuning. Conversely, if short drift lattice adopted, the periodical property of lattice is forced to be break at the interface of the cryomodules and mismatch factor is introduced which leads potential emittance growth within one section.

For avoiding parametric resonance, the zero current phase advances per cell in all the three phase planes are remained below 90° [5]. In the mean while the cavity synchronous phase for each cavity is designed to be bigger than 10 times of the longitudinal Root Mean Square (RMS) beam size to give larger longitudinal acceptance. Due to these limitations, the cavity voltages at the beginning parts of the sections may not be fully exploited, but the cavity voltage should be always bigger than 50% of the nominal design for avoiding multipacting problems in the superconducting spoke cavities. To ensure low loss beam rate, beside the larger longitudinal acceptance, the transverse acceptance criterion is also very important, in this design the beam aperture is kept to be bigger than 8 times of the RMS beam size.

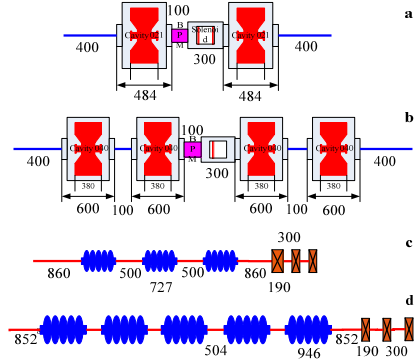

Figure 3: schematic view of the lattice structures for the main linac sections

In order to avoid energy change causing emittance growth and beam quality deterioration by thermal equilibrium between different phase planes, the working points of the linac should be positioned at the resonance-free region in the Hofmann chart [6]. Then even with free energy, no mechanism will drive the energy exchange between different freedoms. In this design the working points are kept to be close to the equipartition line. We fix the zero current phase advance ratio by the emittance ratio according to the approximately equipartitioning condition described in reference [7]:

$$\sigma_{0t}/\sigma_{0z} = k_{t0}/k_{z0} = (\frac{3}{2}\frac{\varepsilon_{nz}}{\varepsilon_{nt}} - \frac{1}{2})^{1/2} \quad (1)$$

where $\sigma_{0t}$, $k_{t0}$ and $\varepsilon_{nt}$ are the transverse zero current phase advance, wave number and normalized emittance, $\sigma_{0z}$, $k_{z0}$ and $\varepsilon_{nz}$ are the longitudinal zero current phase advance、wave number and normalized emittance. Since the phase advance per cell should be below 90°, it is better to set the longitudinal phase advance larger than transverse one to obtain higher acceleration efficiency. This is the reason why in injector Scheme I, the longitudinal emittance out of the RFQ is designed to be smaller than the transverse one.

Between different sections (with different cavity types), there are no extra matching parts added except the intersection between Spoke cavity and Elliptical cavity. The matching is ensured by making the zero current phase advances per meter smooth through varying the parameters of cavities and transverse focusing elements in the adjacent periods. Careful matching should be carried out to avoid important emittance growth at all the transitions. Among the three transitions, the one between the section Spoke040 and the section Ellip063 is the most critical. One reason is the Radio Frequency (RF) frequency doubling from Spoke040 to Ellip063, and the other is the transverse focusing type change from inside-cryomodule solenoids to warm triplet quadrupoles. So that three extra quadrupoles are added between these two sections to help the matching. During the intersection matching, the longitudinal acceptance criterion should be fully kept.

## 3 Main linac design based on injector scheme I

The block diagram of the main linac based on injector scheme I is shown in Figure 4. It is composed of four sections: Spoke021 section, Spoke040 section, Ellip063 section and Ellip082 section.

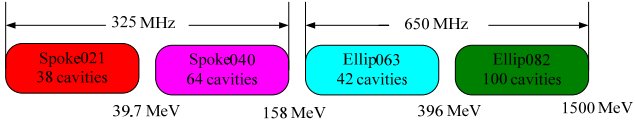

Figure 4: Block diagram of the main linac based on injector scheme I.

### 3.1 General design

The design follows the general design considerations in the last section. Different phase advance ratios between the longitudinal and the transverse planes have been studied. It turns out that the phase advance ratio of 0.75 is adopted after the compromise among the equipartitioning condition, the acceptance to emittance ratio and the phase advances per cell. Figure 5 shows the tune footprint in the Hofmann chart. We can see except one point falling in very weak part of the $k_z/k_x=2$ resonance region, all the other points are in the resonance-free region.

Figure 6 shows the effective RF voltage in use as compared with the nominal voltage for all the four types of superconducting cavities. It is the optimized results by following the requirements on the phase advance, smooth change in focusing and longitudinal acceptance. The voltage ratio is kept bigger than 0.5. To reduce beam loss, the synchronous phase is kept larger than 10 times the RMS phase width throughout the main linac as shown in Figure 7. To be noted, the first two standard spoke021 cells are used for matching together with MEBT2 section. One of the two superconducting spoke cavities in one period is used as a buncher and the other one is a backup for this buncher. The phase advance per meter is shown in Figure 8 and it changes quite smoothly. Basing on this design for main linac, the total cavities number used is 244 and the total length is 385.6m.

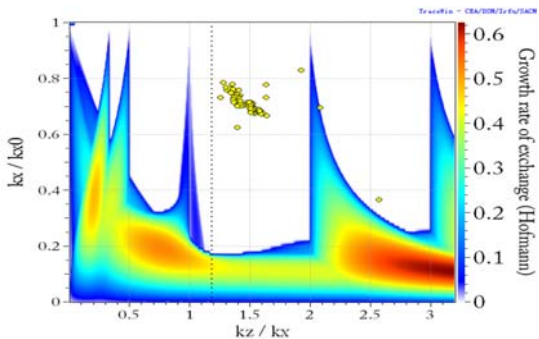

Figure 5: Tune footprint in the Hofmann chart.

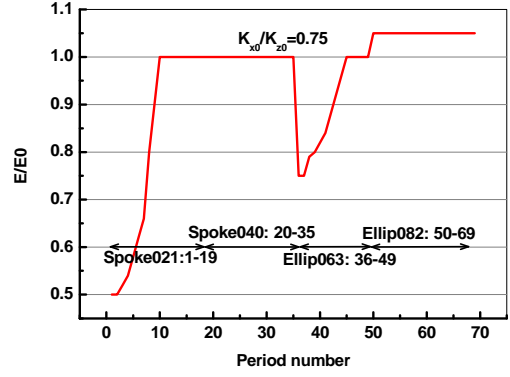

Figure 6: Effective RF voltage in use as compared with the nominal voltage

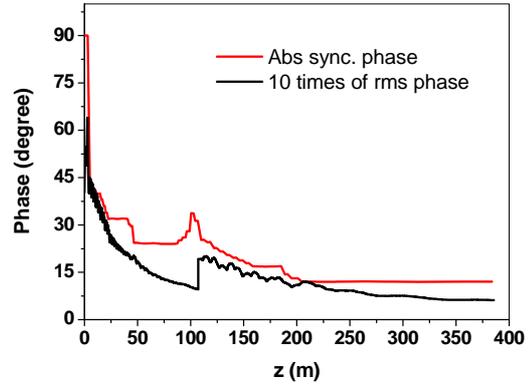

Figure 7: The absolute synchronous phase and 10 times RMS phase width along main linac

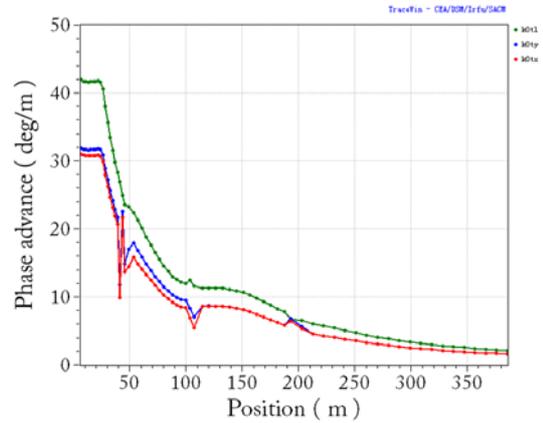

Figure 8: Phase advances per meter along the main linac

### 3.2 Multi-particle simulations

For the multi-particle simulations, it is important to include space charge forces and specify the initial beam distribution. Although most studies use a six dimensional (6D) parabolic distribution, other distributions such as truncated gaussian distribution and simulated distributions at the injectors exit should be also used to check the design robustness. As the first step, we have studied the dynamic behaviour of the beam core and the properties in RMS along the linac. The emittances used for simulations are listed in Table 1. An emittance growth of about 20% is assumed between the exit of RFQ to the entrance of main linac. Without taking into account all kinds of errors and with an input 6D parabolic distribution of $10^5$ macro-particles, the multi-particle simulations for the whole main linac section have been carried out. It is found that the transverse emittance growths are 2.6% and 1.5% for the horizontal and vertical planes, respectively, and the longitudinal emittance growth is 1.6%. The transverse RMS beam size in average is about 2.5 mm. The evolution of the RMS envelope along the main linac is shown in Figure 9. The evolution of the normalized RMS emittance along the main linac is shown in Figure 10. From the simulation results, we can find that the RMS envelope along the main linac is quite smooth. The RMS emittance growth along the main linac is under control in all the three phase spaces, e.g. about a few percent. The tune depressions in the three planes remain as about 0.72 along the linac, and are situated in the transition phase between the space-charge dominant regime and the emittance dominant regime.

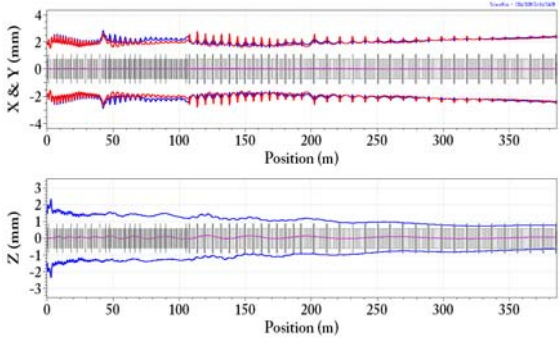

Figure 9: The evolution of the RMS envelope along the main linac

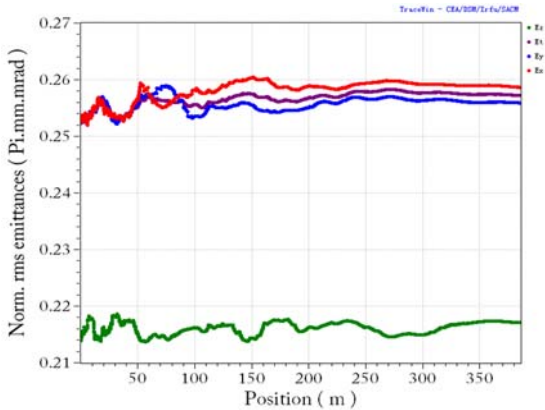

Figure 10: Normalized RMS emittances along the main linac.

Assuming the Injector II output emittance could be the same with Injector I, the main difference for the main linac is the current doubled to be 20 mA. The basic lattice design remains the same, only the matched input Twiss parameters and the matching between the sections are revised. The multi-particle simulation is performed using a 6D parabolic distribution of $10^5$ macro-particles. The simulation results show that the RMS emittance growth along the linac is still under control but clearly larger than that with the beam current of 10 mA, and the envelope evolution is also not as smooth as the 10 mA design.

The error analysis of the design is also performed and details can be found in [3].

### 3.3 Halo formation studies

As the halo development due to errors, mismatches and resonances is the key causing beam loss, it becomes the central focus of the beam dynamics studies once the lattice and the basic dynamic behaviour are determined using the beam core or the RMS emittance. As this is a linac of very high beam power, beam loss should be controlled at the level of $10^{-8}$/m at high energy part. This means that the behaviours of very sparse halo particles should be studied.

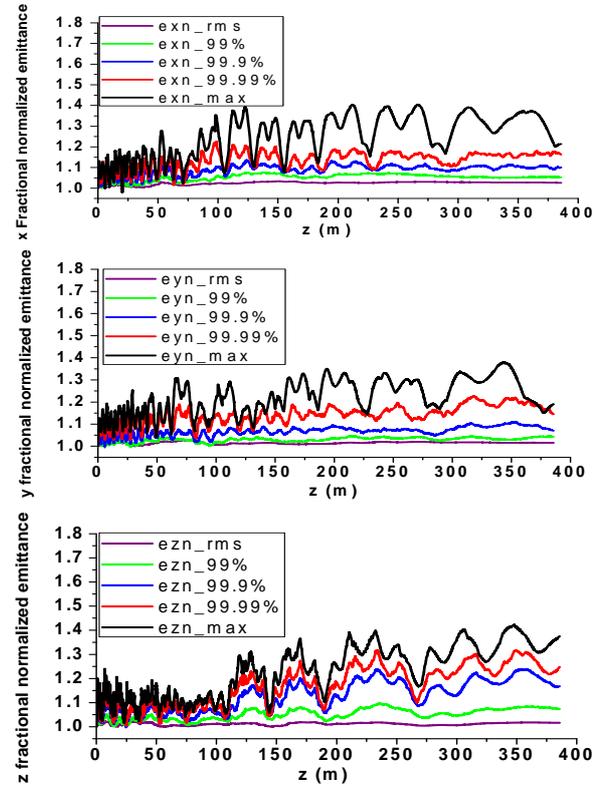

Figure 11: Halo development along the main linac for different beam fractions

For the halo development related to the space charge resonances, we have carried out simulations on the halo formation to see if the working point is sensitive or not. The emittance evolutions for 99%, 99.9%, 99.99% and 100% beam fractions with an input 6D parabolic distribution have been studied, using both TRACK [8] and TraceWin [9] codes (No errors are included in this simulation.). The number of macro-particles is $10^5$ for the simulations. Figure 11 shows the emittance evolutions with the nominal working point ($\varepsilon_z/\varepsilon_x$=0.85, $k_x/k_z$=0.75), which is free from dangerous resonances. The emittance growths with different fractions of particles are below 41%. The result indicates that the basic design is robust.

The halo formation with double bunch current is also studied. It indicates that the maximum emittance growth is about two times the one with 10 mA, which shows the injector scheme of higher RF frequency is favoured from the beam dynamics point of view.

## 4 Main linac design based on injector scheme II

If we examine the previous main linac design with the output beam parameters obtained up to now from the injector scheme II studies, we find the acceptance criteria is violated if the field level and the phase advances criterions are both followed. The main reason is that the longitudinal emittance is much larger in this case. In order to solve this problem, an alternate design modified from the previous design for injector scheme II is proposed.

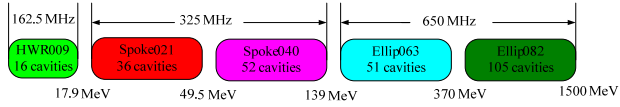

Figure 12: Block diagram of main linac design based on injector scheme II

### 4.1 Base line design

As shown in Figure 12, the main linac on basis of Injector II scheme is composed of five sections. Besides the four sections as used in the previous design, an additional Half Wavelength Resonator (HWR) section with the same cavity type as in the injector II is added in front of Spoke021 section. The lattice structure of the HWR section is similar as the Spoke021 section: each period is composed of two superconducting cavities, one solenoid and one Beam Profile Monitor (BPM). By adding this section, the frequency jump is shifted to higher energy (around 17 MeV), and the longitudinal acceptance condition can be met while keeping the peak field of Spoke021 cavity Ep>12.5MV/m (50% of the nominal design) to avoid multipacting effects. The total cavities number for this design of main linac is 260 and the total length is 414 m. The design results such as synchronous phase, zero current phase advances along the main linac are shown in Figures 13 and 14, respectively. To be noted, the two standard Spoke021 cells in MEBT2 section are not included in this simulation.

### 4.2 Beam dynamics simulations

The beam dynamics programs used for the simulations are TraceWin and Track. An emittance growth of about 20% is assumed between the exit of RFQ to the entrance of main linac. Without taking into account all kinds of errors and with an input 6D parabolic distribution of $10^5$ particles, the multi-particle simulations for the whole main linac have been carried out. As shown in Figure 15, the transverse RMS emittance growths are -1.4% and 6.9% for the horizontal and vertical, respectively, the longitudinal emittance growth is 1.3%. From the simulation results, it is find that the RMS emittance growth along the linac is still under control in all the three phase spaces. Although there is an energy change in the two transverse planes at the beginning of the linac, but they eventually become stable and the emittance growth is not big.

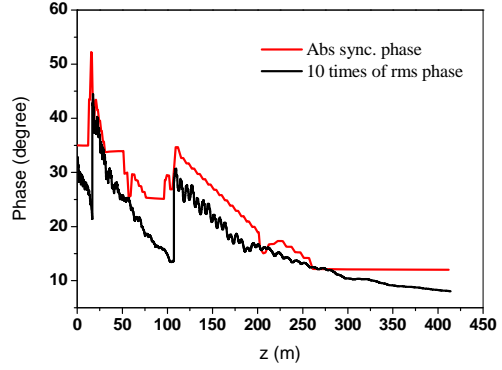

Figure 13: Synchronous phase and 10 times RMS phase width along the main linac with injector scheme II

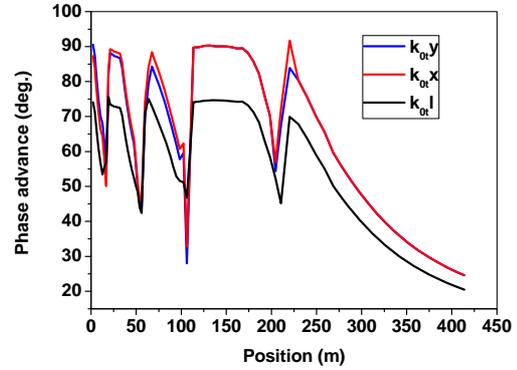

Figure 14: Evolution of transverse (red and blue) and longitudinal (green) zero current phase advances

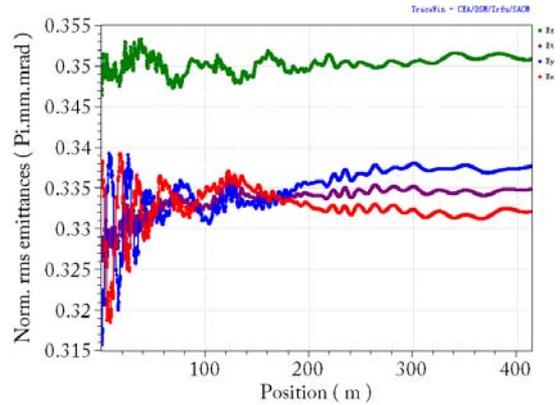

Figure 15: Normalized RMS emittance growths along the main linac with injector scheme II

The beam halo information for this design is also studied. The 100% emittance growths are under 95%, 120% and 70% for horizontal, vertical and longitudinal directions respectively. The relatively larger halo emittance growths are understandable as it is a space charge dominated beam with a bunch current doubled than that with injector I scheme. The envelope evolution is also smooth along the linac and the RMS beam size is about 2.5 mm.

## 5 Fault tolerance capability

In order to ensure the availability of the ADS reactor and avoiding thermal stress causing damage to the subcritical reactor core, the number of unwanted beam trips should not exceed a few per year. This extremely high reliability specification is several orders of magnitude above usual accelerator performance [10]. To decrease beam interruption frequency, the linac has to be designed to handle most of the main elements failure at all stages without loss of the beam. To achieve this requirement, besides all the hardware which is operated with conservative performance and redundancy, it is important to have fault tolerant capabilities in the design [11]. For C-ADS, as described in the earlier section, the injector is guaranteed by a hot stand-by spare, the main linac is guaranteed by means of the local compensation and rematch method.

### 5.1 Local compensation for cavity failure

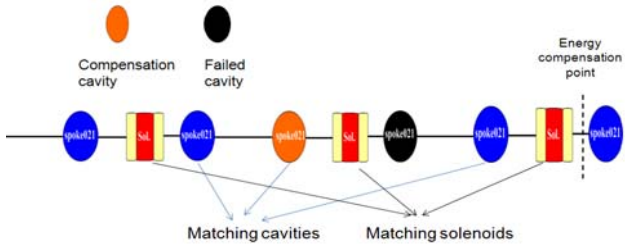

Figure 16: Local compensation scheme of the C-ADS main linac: the first cavity failure in Spoke021 section

For main linac, de-rated cavity gradient is adopted. The nominal peak fields of the spoke cavities are around $E_p$=25 MV/m, and $E_p$=29 MV/m for the elliptical section while the maximum peak field are $E_p$=32.5 MV/m and $E_p$=37 MV/m, respectively (as shown in Table 2). A 30% redundancy is remained for local compensation of the cavities. Once one cavity is failed, the nearest cavities fields are increased (maximum: 1.3 times of the nominal design) to compensate the energy loss, and then the synchronous phases of the surrounding cavities and the solenoids gradients are adjusted to recover the energy and phase at the energy compensation position. The goal is that the "beam trips" could be handled locally without affecting too many devices, so that the energy and time of the beam could be recovered as quickly as possible without affecting the beam transportation.

As an example, the sketch of the local compensation for the first cavity failure in Spoke021 section is shown in figure 16, the black ellipse is standing for a failure cavity and the orange one is presenting cavity for compensation. Cavity failures in three typical positions (in the beginning, end and the middle) in each section are studied and local compensation for total 12 cavities failure in the mean time is simulated. Simulation shows although the halo growth is worse than nominal setting after compensation, but beam is still controllable when there are no errors added. The situation is much better with energy increasing. No beam loss is observed during the simulation with $10^5$ particles 6D Parabolic input distribution.

The main problem is that in this study only energy is compensated, time is not recovered because of the position based program (Tracewin) used in the simulation. Further study for solving this problem is on going, details is reported in reference [3].

### 5.2 Transverse focusing devices failure rematch

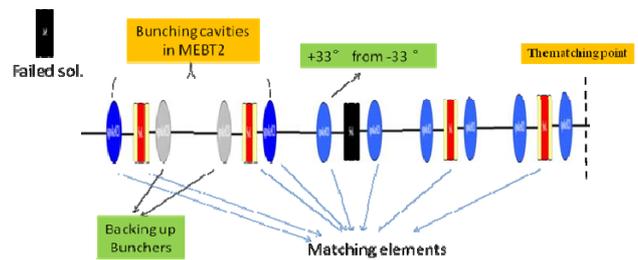

Figure 17: Local compensation scheme of the C-ADS main linac: the first solenoid failure in Spoke021 section

For the solenoid failure, it is found that it is much more difficult to be rematched especially at the low energy section. As we adopted the long period lattice and in one cell there is only one solenoid for transverse focusing, once the solenoid failed, the beam size will become significantly larger even if the beam is rematched by the neighbouring solenoids. Large transverse beam size will result in the emittance distortion due to the nonlinear RF fields. If other elements in the cell could provide additional transverse focusing while the solenoid fails, the beam halo growth could be controlled. Then we studied the possibility of nearby cavity transverse focusing by inversing the synchronous phase of one cavity in the cell. The nominal transverse focusing structure RSR Resonator-Solenoid-Resonator (RSR) or Defocusing-Focusing-Defocusing (DFD) will become FD or DF in the rematch mode. This method has been proven successful.

As focusing element failure is more critical in the cells with large phase advances, namely in the beginning part of each Spoke sections, we take the failure of the first solenoid in the Spoke section as an example. As shown in figure 17, the first solenoid in superconducting Spoke021 section is failed, the synchronous phase of the first cavity in the same period is inversed from -33° to 33°, the neighbour cavities phases and solenoids gradient are adjusted to achieve the local rematching goal at the matching point. The bunching cavities phases are adjusted to recover the time while the energy compensation is achieved. (To be noted, this study is undertaken using an earlier version of main linac design.) Simulation shows after compensation the 100% emittance growth increased up by 10%, 15% and 55% for horizontal, vertical and longitudinal directions, respectively, comparing to the normal gradient. For the solenoid compensation, time compensation still is a problem as cavity is involved in the rematch. Although the time is recovered for this particular case by adjusting the superconducting buncher phases, but this is not feasible for several solenoid failures in the same time.

Usually doublet is widely used for the transverse focusing in higher energy superconducting section, such as Preliminary Design of an eXperimental Accelerator-Driven System (PDS-XADS) project [11] and the Spallation Neutron Source (SNS) at Oak Ridge National Lab [12], but it is found out that, for local compensation, if one quadrupole fails it is better to switch off the whole doublet [13]. Although in this way the beam could be compensated transversely and the RMS emittance growth is also controllable from beam dynamics point of view, but the halo growth could be big while all the errors and mismatch factors are added especially at the beginning of the elliptical section where focusing devices type is changed from solenoids to quadruples, so that the triplets instead of doublets are used in our design for the transverse focusing at the elliptical sections. In case one quadrupole failure, the other two quadrupoles could be used as a doublet. Simulations shows, the halo growth does not appears much difference comparing with the nominal setting after compensation. The energy of this section is much higher could be the main reason. Furthermore, as no cavity is involved in the compensation of quadrupole failure in this section, compensation is much easier.

## 6  Summary and perspective

The C-ADS main linac basic designs and beam dynamics results based on the two injector design schemes have been presented. Longitudinal emittance plays a very important role in designing the lattice structure. The lattices are designed to be conservative to meet the very strict reliability and stability specifications, especially by incorporating the local compensation method. Multi-particle simulations show that the designs are good in controlling emittance and halo growth without taking account of all kinds of errors.

Much more efforts in further optimization of the lattice by including cost trade-off, end-to-end simulations, design robustness with cavity performance variations and different input beam distributions will be needed in the future. Much more efforts will also be needed to the realization of the local compensation strategy.

*The authors express their sincere acknowledgement to the colleagues in the C-ADS accelerator team. Special thanks are expressed to the IMP C-ADS group for providing the Injector II parameters.*